\documentstyle[amsfonts,12pt]{article}
\newcommand{\be}{\begin{equation}}
\newcommand{\ee}{\end{equation}}
\newcommand{\bea}{\begin{eqnarray}}
\newcommand{\eea}{\end{eqnarray}}
\newcommand{\beas}{\begin{eqnarray*}}
\newcommand{\eeas}{\end{eqnarray*}}
\newcommand{\bi}{\begin{itemize}}
\newcommand{\ei}{\end{itemize}}
\newcommand{\bc}{\begin{center}}
\newcommand{\ec}{\end{center}}
\newcommand{\bfl}{\begin{flushleft}}
\newcommand{\efl}{\end{flushleft}}
\newcommand{\bfr}{\begin{flushright}}
\newcommand{\efr}{\end{flushright}}

\def\6{\partial} \def\a{\alpha} \def\b{\beta}

 \def\d{\delta} 

\def\e{\epsilon}

  \def\l{\lambda}

\def\m{\mu} \def\n{\nu}  

 \def\s{\sigma} \def\t{\tau}

\begin{document}

\title{On the $SU\left( 1,1 \right)$ Thermal Group of Bosonic Strings\\
and $D$-Branes}
\author{M. C. B. Abdalla, A. L. Gadelha, I. V. Vancea\\
{\em Instituto de F\'{\i}sica T\'{e}orica, Universidade Estadual Paulista}\\
{\em Rua Pamplona 145, 01405-900 Sao Paulo SP, Brazil}}
\maketitle

\begin{abstract}

All possible Bogoliubov operators that generate the thermal transformations
in the Thermo Field Dynamics (TFD) form a $SU\left( 1, 1 \right)$ group. We 
discuss this construction in the bosonic string theory. In particular, the
transformation of the Fock space and string operators generated by the most
general $SU(1,1)$ unitary Bogoliubov transformation and the entropy of the
corresponding thermal string are computed. Also, we construct the thermal
$D$-brane generated by the $SU(1,1)$ transformation in a constant Kalb-Ramond 
field and compute its entropy. 
\end{abstract}

\setcounter{page}{0}
\thispagestyle{empty}
\newpage

\section{Introduction}

The physical and geometrical properties of $D$-branes have been under
intense investigation for some time. In particular,
the statistics of $D$-branes has attracted a certain attention 
mainly when formulated in the low energy limit of string theory
\cite{mvm,mvmlfb,rab1,rab2,ek,ekwt,kog1,kog2,kog3,od,rsz,gub}. 
Progress has been done in understanding the thermodynamics of black holes and 
string gases 
\cite{as,sv,ls,hp,bfks1,cm,ks,eh,dmkr,hm,ml,bfk,lt,dm,ast,ivvol,ms,mlmjpp,sm}.
Since in this limit the $D$-branes are solitons of (super)gravity, 
the statistics of $D$-branes has been naturally formulated in the Matsubara 
approach to quantum fields at finite temperature. However, despite the success
achieved at low energies, up to now there has been little progress in
understanding the microscopic description and the statistical properties of
$D$-branes in the perturbative limit of string theory where the $D$-branes are
described by coherent states in the Fock space of the closed string sector. 
Due to this interpretation, it is natural do seek for a construction of
thermal branes in the framework of Thermo Field Dynamics in which the 
statistical mechanics is implemented by a thermal Bogoliubov operator
acting on the Fock space and on the operators. 
Along this line of thought, a new approach to thermal $D$-branes has been 
proposed in \cite{ivv,agv,agv1,cei} by using the basic concepts of
Thermo Field Dynamics (TFD) \cite{tu} (which is known to be equivalent to the
Matsubara formalism at thermodynamical equilibrium.)
The construction from \cite{ivv,agv,agv1,cei} has the advantage of maintaining
explicit the interpretation of $D$-branes as coherent boundary states in the 
Fock space of closed bosonic sector. The very interpretation  is used to 
define the $D$-branes at finite temperature 
\cite{clny,pc,gw,bg,gg,ml1,ck,dv1,dv2}.

One of the crucial ingredients of TFD is the thermal Bogoliubov transformation
that maps the 
theory from zero to finite temperature. In particular, one can construct
a thermal vacuum 
$\left.\left| 0 (\b_T )\right\rangle \right\rangle$
annihilated by thermal annihilation operators and one can express the average 
value of any observable $Q$ as the expectation value in the thermal vacuum
\be
Z^{-1}(\b_T ){\mbox Tr}[\rho ~Q]~=~
\left\langle \left\langle 0 (\b_T ) \left| Q \right| 0 (\b_T )\right\rangle 
\right\rangle ,
\label{tfd}
\ee
where $\rho$ is the distribution function, $\b_T ~=~(k_B T)^{-1}$ and $k_B$ is 
the Boltzmann's constant. In order to fulfill the above requirement, the 
vacuum should belong to the direct product space between the 
original Fock space by an identical copy of it denoted by a tilde (see bellow),
which justifies the notation used for vectors. 

As discussed in \cite{um,cu}, 
it is possbile to define various Bogoliubov operators for any theory. They 
form an oscillator representation of $SU(1,1)$ group for bosons and
$SU(2)$ for fermions. If all these operators are taken into account, the
thermal transformation is generated by a linear combination of generators of 
$SU(1,1)$. The coefficients of the generators determine if
the transformation is unitary or not and if the generator satisfy the basic 
requirements of TFD, like the tilde invariance of the most general thermal 
generator and
of vacuum \cite{um,cu}. The aim of this paper is to investigate this group in 
the case of the closed bosonic string and bosonic $D$-brane. This represents
a part of the construction of a TFD-based formulation of thermal bosonic 
vacuum of string theory and thermal $D$-branes.

The outline of the paper is as follows. In the second Section we discuss the 
$SU\left( 1,1 \right)$ group for the case of the bosonic string and construct 
the most general $SU\left( 1,1 \right)$ thermal transformation. The choice of
the parameters in the general Bogoliubov generator is taken such that the
thermal transformation be unitary. With this choice we compute the
entropy of the string. In the third Section we give the boundary conditions 
and the $D$-brane state under the general unitary thermal transformation 
and compute the entropy of the $D$-brane. In the last section we discuss the
results and the main problems raised by implementing the $SU(1,1)$ thermal
group. 

\section{Thermal $SU\left( 1,1 \right)$ Group for Closed String}

Let us consider a closed bosonic string in the Minkowski space-time and in 
the light-cone gauge
$X^{0}\underline{+}X^{25}$. The most general solution of the equations of 
motion with the periodic boundary conditions 
$X^{\mu }\left( \tau ,0\right) =X^{\mu }\left(\tau ,\pi \right) $
has a Fourier expansion. Upon the quantization, the Fourier coefficients are 
interpreted as
operators on the Fock space of the closed string. They must obey the 
canonical commutation relations
\begin{equation}
\left[ \alpha _{k}^{\mu },\alpha _{l}^{\nu }\right] =
k\delta ^{\mu \nu}\delta _{k+l,0},
\label{Fourier}
\end{equation}
where $\alpha _{k}$ denote the operators for the right-hand moving modes. 
Similar 
commution relations hold for the left-hand moving modes $\beta _{k}$ and the 
right-modes and
left-modes are independent \cite{gsw}. It is useful to introduce the 
oscillator operators
\bea
A_{k}^{\mu } &=&\frac{1}{\sqrt{k}}\alpha _{k}^{\mu },\qquad A_{k}^{\mu
\dagger }=\frac{1}{\sqrt{k}}\alpha _{-k}^{\mu },  \nonumber \\
B_{k}^{\mu } &=&\frac{1}{\sqrt{k}}\beta _{k}^{\mu },\qquad B_{k}^{\mu
\dagger }=\frac{1}{\sqrt{k}}\beta _{-k}^{\mu } ~~~\forall~ k~>~0.
\label{oscop}
\eea
Following the TFD construction, we construct the direct product between the 
Fock space of the 
string and an identical copy of it. The quantities referring to the second 
(unphysical) copy of 
string are denoted by $\tilde{~~}$. Thus, the extended space is given by
\begin{equation}
\widehat{{\cal H}}={\cal H}\otimes \widetilde{{\cal H}}.
\label{extspace}
\end{equation}
The vectors of $\widehat{{\cal H}}$ are constructed by acting with the string operators 
on the vacuum state
\begin{equation}
\left. \left| 0\right\rangle \!\right\rangle =\left. \left| 0\right\rangle
\!\right\rangle _{\alpha }\left. \left| 0\right\rangle \!\right\rangle
_{\beta }=\left( \left| 0\right\rangle \!_{\alpha }\widetilde{\left|
0\right\rangle }\!_{\alpha }\right) \left( \left| 0\right\rangle \!_{\beta }
\widetilde{\left| 0\right\rangle }\!_{\beta }\right) =\left( \left|
0\right\rangle \!_{\alpha }\left| 0\right\rangle \!_{\beta }\right) \left( 
\widetilde{\left| 0\right\rangle }\!_{\alpha }\widetilde{\left|
0\right\rangle }\!_{\beta }\right),
\label{vacuum}
\end{equation}
where the vacua of the right/left Fock spaces are given by
\begin{eqnarray}
\left. \left| 0\right\rangle \!\right\rangle _{\alpha } &=&\left|
0\right\rangle \!_{\alpha }\otimes \widetilde{\left| 0\right\rangle }
_{\alpha }=\left| 0,0\right\rangle _{\alpha }\!,  \nonumber \\
\left. \left| 0\right\rangle \!\right\rangle _{\beta } &=&\left|
0\right\rangle _{\beta }\otimes \!\widetilde{\left| 0\right\rangle }_{\beta
}\!=\left| 0,0\right\rangle \!_{\beta }.
\label{rightleft}
\end{eqnarray}
The original string and its copy with each other. Therefore, the operator 
algebra 
of the extended system is given by the following relations
\begin{eqnarray}
\left[ A_{k}^{\mu },A_{m}^{\nu \dagger }\right]  &=&\left[ \tilde{A}
_{k}^{\mu },\tilde{A}_{m}^{\nu \dagger }\right] =\delta _{km}\eta ^{\mu \nu
},  \nonumber \\
\left[ A_{k}^{\mu },\tilde{A}_{m}^{\nu }\right]  &=&\left[ A_{k}^{\mu },
\tilde{A}_{m}^{\nu \dagger }\right] =\left[ A_{k}^{\mu },\tilde{B}_{m}^{\nu
}\right] =...=0.
\label{opalg}
\end{eqnarray}
The Hamiltonian of the extended system is constructed by demanding that the 
thermal vacuum, which will be constructed later, be invariant under time 
translation.
This can be obtained if we define the following Hamiltonian operator
\be
\hat{H} ~=~H-\tilde{H}  
~=~\sum_{n >0}^{\infty }
n\left( A_{n}^{\dagger }\cdot A_{n}+ B_{n}^{\dagger }\cdot B_{n}
 -  \tilde{A}_{n}^{\dagger }\cdot \tilde{A}_{n}- 
\tilde{B}_{n}^{\dagger }\cdot \tilde{B}_{n}\right) 
\label{exthamilt}
\ee
The relation (\ref{exthamilt}) shows that, if the new vacuum state is time 
invariant, then
the copy of the string is non-physical. Indeed, by adding the second Fock 
space corresponds 
to furnishing new thermal degrees of freedom rather than dynamical degrees of 
freedom, necessary
for discussing the statistics. Consequently, the physical observables are 
defined by operators
without tilde.

In order to construct a finite temperature model of string, one has to act on 
zero temperature
Fock space and oscillator operators with the Bogoliubov operators which 
produce a thermal noise
\cite{um}.

\subsection{Thermal Bogoliubov Transformation}
 
The first step in constructing the string theory at finite temperature 
is to provide a thermal vacuum.
This can be obtained by acting on the extended system vacuum 
(\ref{vacuum}) with any operator that: i) mixes the operators $A_{k}^{\mu }$, 
$\tilde{A}_{k}^{\mu \dagger }$ for the right modes and $B_{k}^{\mu }$, 
$\tilde{B}_{k}^{\mu \dagger }$ for the left modes; ii) commutes with the 
Hamiltonian (\ref{exthamilt}) and 
iii) takes the form of a Bogoliubov transformation \cite{um}. 
The right/left transformations should be independent from the algebra 
(\ref{opalg}).  The general form of the Bogoliubov transformation that fix 
the form of the generator is given by the following relation \cite{um,cu}
\begin{equation}
\left( 
\begin{array}{c}
A^{\prime } \\ 
\tilde{A}^{\dagger \prime }
\end{array}
\right) =e^{-iG}\left( 
\begin{array}{c}
A \\ 
\tilde{A}^{\dagger }
\end{array}
\right) e^{iG}={\cal B}\left( 
\begin{array}{c}
A \\ 
\tilde{A}^{\dagger }
\end{array}
\right) ,\quad \left( 
\begin{array}{cc}
A^{\dagger ^{\prime }} & -\tilde{A}^{\prime }
\end{array}
\right) =\left( 
\begin{array}{cc}
A^{\dagger } & -\tilde{A}
\end{array}
\right) {\cal B}^{-1},
\label{Bogoliubov}
\end{equation}
where $\cal B $ is a $2 \times 2$ complex unitary matrix 
 \begin{equation}
{\cal B}=\left( 
\begin{array}{cc}
u & v \\ 
v^{*} & u^{*}
\end{array}
\right) ,\qquad \left| u\right| ^{2}-\left| v\right| ^{2}=1,
\label{Bmatrix}
\end{equation}
and $G$ is the generator of the transformation, called 
{\em the Bogoliubov operator}. The operators
that satisfy the relations (\ref{Bogoliubov}) and (\ref{Bmatrix}) have the 
following form \cite{cu}
\begin{eqnarray}
g_{1_{k}}^{\alpha } &=&\theta _{1_{k}}\left( A_{k}\cdot \tilde{A}_{k}+\tilde{
A}_{k}^{\dagger }\cdot A_{k}^{\dagger }\right) , ~~~~~\qquad 
g_{1_{k}}^{\beta
}=\theta _{1_{k}}\left( B_{k}\cdot \tilde{B}_{k}+\tilde{B}_{k}^{\dagger
}\cdot B_{k}^{\dagger }\right) ,  \nonumber \\
g_{2_{k}}^{\alpha } &=&i\theta _{2_{k}}\left( A_{k}\cdot \tilde{A}_{k}-
\tilde{A}_{k}^{\dagger }\cdot A_{k}^{\dagger }\right) ,~~~~\qquad
g_{2_{k}}^{\beta }=i\theta _{2_{k}}\left( B_{k}\cdot \tilde{B}_{k}-\tilde{B}
_{k}^{\dagger }\cdot B_{k}^{\dagger }\right) ,  \nonumber \\
g_{3_{k}}^{\alpha } &=&\theta _{3_{k}}\left( A_{k}^{\dagger }\cdot A_{k}+
\tilde{A}_{k}^{\dagger }\cdot \tilde{A}_{k}+\delta_{kk}tr\eta ^{\mu \nu} \right),\qquad
g_{3_{k}}^{\beta }=\theta _{3_{k}}\left( B_{k}^{\dagger }\cdot B_{k}+\tilde{B
}_{k}^{\dagger }\cdot \tilde{B}_{k}+\delta_{kk}tr\eta ^{\mu \nu} \right) ,
\label{generators}
\end{eqnarray}
where the index $\a$ refeers to the right-moving modes and $\b$ to the left-moving modes,
respectively and the $\theta$'s are real parameters depending on the 
temperature which, for convenience, have been included in the operators. 
We assume that they are  are monotonous  increasing functions 
on $T$.
It is easy to very that the generators (\ref{generators}) satisfy the 
$SU\left( 1,1 \right)$ algebra
\be
\left[ g_{1_{k}}^{\alpha ,\beta },g_{2_{k}}^{\alpha ,\beta }\right]
=-i\Theta _{123}g_{3_{k}}^{\alpha ,\beta },\quad \left[ g_{2_{k}}^{\alpha
,\beta },g_{3_{k}}^{\alpha ,\beta }\right] =i\Theta _{231}g_{1_{k}}^{\alpha
,\beta },\quad \left[ g_{3_{k}}^{\alpha ,\beta },g_{1_{k}}^{\alpha ,\beta
}\right] =i\Theta _{312}g_{2_{k}}^{\alpha ,\beta },
\label{su11}
\ee
where we have defined
\begin{equation}
\Theta _{ijk}\equiv 2\frac{\theta _{i_{k}}\theta _{j_{k}}}{\theta _{k_{k}}}.
\label{thetas}
\end{equation}
As we can see from  (\ref{generators}), the most general thermal 
transformation takes the following form
\begin{equation}
G=\sum_{k}\left( G_{k}^{\alpha }+G_{k}^{\beta }\right) ,
\label{gentransf}
\end{equation}
where the right/left generators are given by the relations
\begin{eqnarray}
G_{k}^{\alpha } &=&\lambda _{1_{k}}\tilde{A}_{k}^{\dagger }\cdot
A_{k}^{\dagger }-\lambda _{2_{k}}A_{k}\cdot \tilde{A}_{k}+\lambda
_{3_{k}}\left( A_{k}^{\dagger }\cdot A_{k}+\tilde{A}_{k}^{\dagger }\cdot 
\tilde{A}_{k}+\delta_{kk}tr\eta ^{\mu \nu} \right) , \\
G_{k}^{\beta } &=&\lambda _{1_{k}}\tilde{B}_{k}^{\dagger }\cdot
B_{k}^{\dagger }-\lambda _{2_{k}}B_{k}\cdot \tilde{B}_{k}+\lambda
_{3_{k}}\left( B_{k}^{\dagger }\cdot B_{k}+\tilde{B}_{k}^{\dagger }\cdot 
\tilde{B}_{k}+\delta_{kk}tr\eta ^{\mu \nu} \right)
\label{rlgen}
\end{eqnarray}
and the coefficients represent complex linear combinations of $\theta$'s
\begin{equation}
\lambda _{1_{k}}=\theta _{1_{k}}-i\theta _{2_{k}},\quad \lambda
_{2_{k}}=-\lambda _{1_{k}}^{*},\quad \lambda _{3_{k}}=\theta _{3_{k}}.
\label{lambdas}
\end{equation}
The operator (\ref{gentransf}) generates the thermal transformation. The 
dependence on temperature is contained in the complex parameters $\lambda$. 

There is some arbitrariety in choosing the parameters $\theta$. This freedom 
can be used to fix the type of transformation. There are two conditions that
are imposed on a general thermal transformation: i) unitarity and ii) 
invariance under tilde transformation that acts on any arbitrary operator as
follows
\be
(AB)^{\sim} = \tilde{A}\tilde{B}~~~,~~~
(\alpha A)^{\sim} = \alpha^{*}\tilde{A},
\label{tildeop}
\ee
where $\alpha$ is a complex number and ${*}$ is the complex conjugation. The
invariance under tilde operation guarantees the invariance of the thermal 
vacuum under the same operation. However, working with the thermal $SU(1,1)$ 
implies choosing one type of transformation. The unitarity and
the tilde invariance are not always simultaneously compatible \cite{um,cu}. 
In general, the two conditions will select only the generator $g_{2_{k}}$
from the three generators above and reduce the problem to the TFD with one
generator. In what follows, we will stick to the unitarity condition as being
the most natural one for the system at hand, i. e. string theory. The other
choice will be commented in the final section.

\subsection{Thermal Vacuum and Thermal String Operators}

The thermal vacuum of the system at finite temperature is obtained 
by acting on the 
vacuum at zero temperature (\ref{vacuum}) with the operator (\ref{gentransf}) 
\cite{tu}
\begin{equation}
\left. \left| 0\left( \theta \right) \right\rangle \!\right\rangle
=e^{-iG}\left. \left| 0\right\rangle \!\right\rangle .
\label{thermvac}
\end{equation}
Since the left/right-moving terms commute among themselves, there are two 
distinct 
contributions from the
$\a$ sector and the $\b$ sector, respectively. Consider the right-hand 
thermal vacuum. 
By applying the 
disentanglement theorem for $SU \left( 1,1 \right)$ \cite{cha,eber}, one can 
write the thermal 
vacuum under the following form
\bea
\left. \left| 0\left( \theta \right) \right\rangle \!\right\rangle
=\prod_{k}e^{\Gamma _{1_{k}}\left( \tilde{A}_{k}^{\dagger }\cdot
A_{k}^{\dagger }\right) }e^{\log \left( \Gamma _{3_{k}}\right) \left(
A_{k}^{\dagger }\cdot A_{k}+\tilde{A}_{k}^{\dagger }\cdot \tilde{A}
_{k}+\delta_{kk}tr\eta ^{\mu \nu}\right) }e^{\Gamma _{2_{k}}
\left( A_{k}\cdot \tilde{A}_{k}\right)}\\
\times e^{\Gamma _{1_{k}}\left( \tilde{B}_{k}^{\dagger }\cdot
B_{k}^{\dagger }\right) }e^{\log \left( \Gamma _{3_{k}}\right) \left(
B_{k}^{\dagger }\cdot B_{k}+\tilde{B}_{k}^{\dagger }\cdot \tilde{B}
_{k}+\delta_{kk}tr\eta ^{\mu \nu}\right) }e^{\Gamma _{2_{k}}
\left( B_{k}\cdot \tilde{B}_{k}\right)}
\left. \left| 0\right\rangle \!\right\rangle  \label{vacalpha}
\eea
where the coefficients of various generators are given by the relations
\be
\Gamma _{1_{k}}=\frac{-\lambda _{1_{k}}\sinh \left( i\Lambda _{k}\right) }{%
\Lambda _{k}\cosh \left( i\Lambda _{k}\right) +\lambda _{3_{k}}\sinh \left(
i\Lambda _{k}\right) },\quad \Gamma _{2_{k}}=\frac{\lambda _{2_{k}}\sinh
\left( i\Lambda _{k}\right) }{\Lambda _{k}\cosh \left( i\Lambda _{k}\right)
+\lambda _{3_{k}}\sinh \left( i\Lambda _{k}\right) },
\label{lambda12}
\ee
\begin{equation}
\Gamma _{3_{k}}=\frac{\Lambda _{k}}{\Lambda _{k}\cosh \left( i\Lambda
_{k}\right) +\lambda _{3_{k}}\sinh \left( i\Lambda _{k}\right) },
\label{lambda3}
\end{equation}
and 
\begin{equation}
\Lambda _{k}^{2}\equiv \left( \lambda _{3_{k}}^{2}+\lambda _{1_{k}}\lambda
_{2_{k}}\right) .
\label{biglambda}
\end{equation}
Since the vacuum at zero temeprature is annihilated by $A_{k}^{\mu }$
and $\tilde{A}_{k}^{\mu }$,
the only contribution to the thermal vacuum is given by
\begin{equation}
\left. \left| 0(\theta )\right\rangle \!\right\rangle
=\prod_{k}(\Gamma _{3_{k}})^{2\delta_{kk}tr\eta ^{\mu \nu}}
e^{\Gamma _{1_{k}}\left( \tilde{A}_{k}^{\dagger
}\cdot A_{k}^{\dagger }\right) }e^{\Gamma _{1_{k}}\left( \tilde{B}_{k}^{\dagger
}\cdot B_{k}^{\dagger }\right) }\left. \left| 0\right\rangle \!\right\rangle.
\label{thermvacfin}
\end{equation}
The thermal vacuum of the left-moving modes is constructed in the same way. 
The total vacuum at 
finite temperature is the direct product between the $\a$ and $\b$ vacua. The
string operators
are mapped to finite temperature by the corresponding Bogoliubov generators
\begin{eqnarray}
A_{k}^{\mu }\left( \theta \right)  &=&e^{-iG_{k}^{\alpha }}A_{k}^{\mu
}e^{iG_{k}^{\alpha }},\qquad \tilde{A}_{k}^{\mu }\left( \theta \right)
=e^{-iG_{k}^{\alpha }}\tilde{A}_{k}^{\mu }e^{iG_{k}^{\alpha }},  \nonumber \\
B_{k}^{\mu }\left( \theta \right)  &=&e^{-iG_{k}^{\beta }}B_{k}^{\mu
}e^{iG_{k}^{\beta }}~~\qquad \tilde{B}_{k}^{\mu }\left( \theta \right)
=e^{-iG_{k}^{\beta }}\tilde{B}_{k}^{\mu }e^{iG_{k}^{\beta }}.
\label{thermop}
\end{eqnarray}
Similar relations hold for the creation operators. One can easily show that 
the thermal
operators satisfy the same canonical commutation relations as the operators at
zero temperature.
Alternatively, one can organize the operators in thermal doublets \cite{um,cu} 
and
obtain the thermal operators by acting on the doublet with $\cal B$ matrix 
\begin{equation}
\left( 
\begin{array}{c}
A_{k}^{\mu }\left( \theta \right)  \\ 
\tilde{A}_{k}^{\mu \dagger }\left( \theta \right) 
\end{array}
\right) ={\cal B}_{k}\left( 
\begin{array}{c}
A_{k}^{\mu } \\ 
\tilde{A}_{k}^{\mu \dagger }
\end{array}
\right) ,
\label{doublet}
\end{equation}
where the explicit form of ${\cal B}_k$ operators is given by the following 
relation
\begin{equation}
{\cal B}_{k}=\cosh \left( i\Lambda _{k}\right) {\Bbb I} +\frac{\sinh \left(
i\Lambda _{k}\right) }{\left( i\Lambda _{k}\right) }\left( 
\begin{array}{cc}
i\lambda _{3_{k}} & i\lambda _{1_{k}} \\ 
i\lambda _{2_{k}} & -i\lambda _{3_{k}},
\end{array}
\right) 
\label{explB}
\end{equation}
where $\Bbb I$ is the identity matrix. 
With all these elements at hand, one can construct a solution of the string 
equations of motion
with periodic boundary conditions at finite temperature. This reduces to 
replacing the string
operators by thermal string operators. Since the later satisfy the usual 
canonical commutations,
the equations of motion and its solution satisfy the Virasoro algebra with the
following thermal generators
\begin{equation}
L_{m}^{\alpha }\left( \theta \right) =\frac{1}{2}\sum_{k\in {\cal Z}}\alpha
_{-k}\left( \theta \right) \alpha _{k+m}\left( \theta \right) ,\quad
L_{m}^{\beta }\left( \theta \right) =\frac{1}{2}\sum_{k\in {\cal Z}}\beta
_{-k}\left( \theta \right) \beta _{k+m}\left( \theta \right) 
\label{thermalVir}
\end{equation}
which guarantees that we are working with thermal strings \cite{ivv,agv}.  

\subsection{Entropy of Thermal String}

The entropy operator is defined such that its average value be proportional to
the
entropy of the bosonic field at thermal equilibrium divided by the Boltzmann's
constant 
\cite{tu}. From (\ref{tfd}), the entropy of the bosonic field can be computed
as the 
expectation value of the entropy operator in the thermal vacuum
\begin{equation}
\frac{1}{k_{B}}\left\langle \!\left\langle 0\left( \theta \right) \right|
\right. K\left. \left| 0\left( \theta \right) \right\rangle \!\right\rangle
=\left\{ \sum_{k}\left[ \left( 1+n_{k}\right) \log \left( 1+n_{k}\right)
-n_{k}\log \left( n_{k}\right) \right] \right\},
\label{entrboson}
\end{equation}
where $n_k$ is the density of particles. Consequently, one defines the 
entropy operator for the bosonic string as
\begin{equation}
K=K^{\alpha }+K^{\beta },
\label{entrstring}
\end{equation}
where the entropies of the right- and left-moving modes are given by the 
following relations
\bea
K^{\alpha } &=& -\sum_{k}\left[ A_{k}^{\dagger }\cdot A_{k}\log \left( g\frac{
\lambda _{1_{k}}\lambda _{2_{k}}}{\Lambda _{k}^{2}}\sinh ^{2}\left( i\Lambda
_{k}\right) \right) -A_{k}\cdot A_{k}^{\dagger }\log \left( 1+g\frac{\lambda
_{1_{k}}\lambda _{2_{k}}}{\Lambda _{k}^{2}}\sinh ^{2}\left( i\Lambda
_{k}\right) \right) \right] \label{rightentr}\\
K^{\beta }&=&-\sum_{k}\left[ B_{k}^{\dagger }\cdot B_{k}\log \left( g\frac{
\lambda _{1_{k}}\lambda _{2_{k}}}{\Lambda _{k}^{2}}\sinh ^{2}\left( i\Lambda
_{k}\right) \right) -B_{k}\cdot B_{k}^{\dagger }\log \left( 1+g\frac{\lambda
_{1_{k}}\lambda _{2_{k}}}{\Lambda _{k}^{2}}\sinh ^{2}\left( i\Lambda_{k}
\right) \right) \right] .
\label{leftentr}
\eea
With this definition of the entropy operator one recovers the entropy of 
\cite{agv} in the
case when there is a single transformation generated by $g_{2_k}$, that is
when $\theta _{1_{k}}=\theta _{3_{\kappa }}=0$. Also, this choice gives the 
operator of \cite{tu} when $g = 1$. One obtains the entropy of the bosonic
closed string as the average of (\ref{entrstring}) in the thermal vacuum
\begin{eqnarray}
S &=&k_{B}\left\langle \!\left\langle 0\left( \theta \right) \right| \right.
K\left. \left| 0\left( \theta \right) \right\rangle \!\right\rangle  
\nonumber \\
&=&2k_{B}\sum_{k}\left[ \left( g+n_{k}\right) \log \left( 1+n_{k}\right)
-n_{k}\log \left( n_{k}\right)\right],
\label{closedstringentr} 
\end{eqnarray}
where 
\begin{equation}
n_{k}=g\left[ \frac{\lambda _{1_{k}}\lambda _{2_{k}}}{\Lambda _{k}^{2}}
\sinh^{2}\left( i\Lambda _{k}\right) \right] ,
\label{nnumb}
\end{equation}
and
\begin{equation}
g=\left\langle \!\left\langle 0\right| \right. \tilde{A}_{k}\cdot \tilde{A}
_{k}^{\dagger }\left. \left|0\right\rangle \!\right\rangle =
\left\langle \!\left\langle 0\right| \right. \tilde{B}_{k}\cdot
\tilde{B}_{k}^{\dagger }\left. \left|0\right\rangle \!\right\rangle .
\label{gaver}
\end{equation}
Note that the entropy operator for the bosonic closed string has been 
constructed as the sum between
the right- and the left-moving modes treated as two independent subsystems of
 the string. Thus, the
entropy operator obeys the extensivity property of the physical quantity. 
Also, from 
(\ref{closedstringentr}), we see that entropy of the bosonic closes string 
goes to zero when
the system is in equilibrium, that is  
\begin{equation}
n_{k}=\frac{e^{-\left( k_{B}T\right) ^{-1}\omega }}{1-e^{-\left(
k_{B}T\right) ^{-1}\omega }},
\label{nequi}
\end{equation}
and  take the limit $T\rightarrow 0$. This guarantees that third principle of 
the Thermodynamics is satisfied \cite{hua}. 
 
\section{Boundary States under $SU\left( 1, 1 \right)$ Transformation}

At $T =0$, a rigid $Dp$-brane located along the $\{ X^a \}$ directions in 
the target-space, $a=1,2,\ldots,p$ at $\{ X^i = y^i \}$, where 
$i=p+1,\ldots 24$ in
the presence of the constant Kalb-Ramond field $F_{ab}$. The $Dp$-brane is
described by a superposition of coherent boundary states in the Fock space
defined by the boundary conditions in the closed string sector. In order to 
find the thermal $Dp$-brane, one has to write down the boundary conditions at
finite temperature. This can be achieved by interpreting the string coordintes
and their derivatives as operators and then by acting on them with the 
Bogoliubov transformation. In the case when one considers
just a single Bogoliubov generator \cite{ivv,agv}, the boundary conditions at
finite 
temperature admit solution. In this section we are going to investigate the 
case when the thermal transformation is generated by the general 
$SU \left( 1,1 \right)$ 
generator, i.e. by the Bogoliubov operator (\ref{gentransf}). 

\subsection{Thermal Boundary Conditions and Boundary States}

The boundary conditions in the closed string sector at $T=0$ are given by the 
following relations
\begin{eqnarray}
\left. \left( \partial _{\tau }X_{a}\left( \t , \s \right) +F_{ba}\partial
_{\sigma }X^{b}\left( \t , \s \right) \right) \right| _{\tau =0} &=&0,\\
\left. X^{i}\left( \t , \s \right)-y^{i} \right| _{\tau =0}&=&0.
\label{boundcond}
\end{eqnarray}
One can obtain the corresponding relations for the second copy of the string 
from 
(\ref{boundcond}) by replacing the string coordinates by the coordinates of 
the tilde string.
In order to find the boundary conditions at finite temperature, one 
substitutes the
general solution of the equations of motion in (\ref{boundcond}) and apply the
general Bogoliubov (\ref{gentransf})
\be
X^{\m}(\theta) = e^{-iG}X^{\m}e^{iG}.
\label{stringbogol}
\ee 
The thermal boundary conditions in terms of string operators at finite 
temperature 
represent a set of constraints on the Fock space of the extended system of 
the following form \cite{ivv,agv}
\begin{eqnarray}
 \left[ \left( {\Bbb I}+\hat {{\Bbb F}}\right) _{b}^{a}A_{n}^{b}\left( \theta \right)
+\left( {\Bbb I}+\hat {{\Bbb F}}\right) _{b}^{a}  B_{n}^{b\dagger }\left( \theta
\right) \right] \left.\left| B(\theta ) \right\rangle \! \right\rangle  &=&0, 
\nonumber \\
\left[ \left( {\Bbb I}+\hat {{\Bbb F}}\right) _{b}^{a}A_{n}^{b\dagger }\left( \theta
\right) +\left( {\Bbb I}+\hat {{\Bbb F}}\right) _{b}^{a}B_{n}^{b}\left( \theta
\right)\right] \left.\left| B(\theta ) \right\rangle \! \right\rangle   &=&0, 
\nonumber \\
\left[ A_{n}^{i}\left( \theta \right) -B_{n}^{i\dagger }
\left( \theta \right)\right] 
\left.\left| B(\theta ) \right\rangle \! \right\rangle  &=&0, \nonumber \\
\left[A_{n}^{i\dagger }\left( \theta \right) -B_{n}^{i}\left( \theta 
\right)\right] 
\left.\left| B(\theta ) \right\rangle \! \right\rangle  &=&0,
\label{boundarycondT}
\eea
for any $n>0$. The coordinates of the center of mass and their conjugate 
momenta 
do not transform under the Bogoliubov transformation, neither does the 
constant Kalb-Ramond field \cite{ivv,agv}. Then, we have to add to the set 
(\ref{boundarycondT}) the following
relations
\be
\hat{p}^{a} \left.\left| B(\theta ) \right\rangle \! \right\rangle =
\left[ \hat{q}^{i}-y^{i} \right] \left.\left| B(\theta ) \right\rangle 
\! \right\rangle  =0.
\label{condmomT}
\ee
Similar boundary conditions should be imposed for the tilde string and the 
thermal 
$Dp$-brane $\left.\left| B(\theta ) \right\rangle \right\rangle $ must 
satisfy the 
two of them. It is easy to see that a general solution of the equations 
(\ref{boundarycondT}) and (\ref{condmomT}) has the following form
\bea
\left. \left| B\left( \theta \right) \right\rangle \! \right\rangle _{1} 
&=&N_{p}^{2}\left( F,\theta \right) \delta ^{\left( d_{\perp}\right) }
\left( \hat{q}-y\right) \delta ^{\left( d_{\perp }\right) }\left( 
\widetilde{\hat{q}}-\widetilde{y}\right) \times \nonumber\\   
&&e^{-\sum\limits_{n=1}^{\infty }
A_{n}^{\dagger }
\left( \theta \right) 
\cdot M\cdot
B_{n}^{\dagger }
\left( \theta \right)}
e^{-\sum\limits_{n=1}^{\infty }
\widetilde{A}_{n}^{\dagger }
\left( \theta \right) 
\cdot M\cdot 
\widetilde{B}_{n}^{\dagger }
\left( \theta \right) } \left. 
\left| 0\left( \theta \right) \right\rangle \! \right\rangle, 
\label{solT1}
\eea
where 
\begin{equation}
M_{\nu }^{\mu }=\left[ \left( \frac{{\Bbb I}-\hat {{\Bbb F}}}{{\Bbb I}+\hat {{\Bbb F}}}
\right) _{b}^{a};-\delta _{j}^{i}\right]
\label{Mmatrix}
\end{equation}
and
$N_p(F,\theta)$ is the thermal normalization constant equal to the one of the 
tilde
system and identical to the normalization constant at $T = 0$ \cite{agv}
\be
N_{p}\left( F,\theta \right) = N_p \left( F \right) = 
\sqrt{- \mbox{det}(\delta + 2\pi \a 'F)}.
\label{normconst}
\ee
Note that (\ref{normconst}) corresponds to the first solution found in 
\cite{agv}. We 
interpret this solution as describing a thermal $Dp$-brane and postpone to 
the last section the discussion of the degeneracy of thermal brane in this 
case.

\subsection{The Entropy of Thermal $D$-Branes}

The thermal $D$-brane given by (\ref{solT1}) represent a superposition  of 
coherent states in the Fock space of the thermal string. Therefore,  
calculating the entropy of the $D$-brane is equivalent to computing the 
average value of the entropy operator 
({\ref{entrstring}) in the state (\ref{solT1}). One way of doing that is by 
expressing all the operators and states in terms of operators and states at 
$T \neq 0$. To this end we need the inverse of the Bogoliubov matrix 
(\ref{explB}) which has the following form
\begin{equation}
{\cal B}^{-1}_{k}=  \cosh \left( i\Lambda _{k}\right) {\Bbb I} -
\frac{\sinh \left(
i\Lambda _{k}\right) }{\left( i\Lambda _{k}\right) }\left( 
\begin{array}{cc}
i\lambda _{3_{k}} & i\lambda _{1_{k}} \\ 
i\lambda _{2_{k}} & -i\lambda _{3_{k}}
\end{array}
\right), 
\label{invB}
\end{equation}
By using (\ref{invB}), we can write the entropy operator $K^{\a}$ in terms of
the thermal operators
\bea
A^\mu_k &=& \left[\cosh(i\Lambda_k) - 
\frac{\sinh(i\Lambda_k)}{\Lambda_k}\lambda_{3k}\right]
A_k^\mu(\theta) - 
\frac{\sinh(i\Lambda_k)}{\Lambda_k}\lambda_{1k}\tilde A^{\mu\dagger}_k(\theta)
\label{thermA}\nonumber\\
A^{\mu\dagger}_k &=& \left[\cosh(i\Lambda_k) - 
\frac{\sinh(i\Lambda_k)}{\Lambda_k}
\lambda_{3k}\right]A_k^{\mu\dagger}(\theta) - 
\frac{\sinh(i\Lambda_k)}{\Lambda_k}\lambda_{2k}\tilde A^{\mu}_k(\theta).
\label{thermtildeA}
\nonumber\\
\eea
Then, the entropy in the right-moving sector has the following form 
\be 
\left\langle \!\left\langle 0\left( \theta \right) \right| \right.
K\left. \left| 0\left( \theta \right) \right\rangle \!\right\rangle =
- \sum_{k} \{ \left [1+2n_k \right] {\cal A}_k \log \left( \frac{n_k}{1+n_k} \right)
+ 
n_k\log \left( {n_k} \right)
-\left( g+n_k \right) \log(1+n_k) \},
\label{rightentro}  
\ee
where
\be
n_k = g \frac{\l_{1k}\l_{2k}}{\Lambda_{k}}\sinh^2(i\Lambda_k),
\label{Tm}
\ee
as in (\ref{nnumb}), and
\be
{\cal A}_k = \left\langle\!\!\!\left\langle B(\theta)\Bigg\vert 
\sum_{\mu=1}^{24} N_k^\mu (\theta )
\Bigg\vert B (\theta)\right\rangle\!\!\!\right\rangle .
\label{Am}
\ee
In order to calculate the action of the thermal number operator, we 
expand the exponentials in the coherent state. Since all the operators are
at non-zero temperature, we ignore the $\theta$ simbol in the notation. For
example, by expanding the non-tilde part of the coherent state we obtain the 
following expression for the ${\cal A}_k$
\be
B^2\tilde B^2\left\langle\!\!\!\left\langle 0(\theta)
\Bigg\vert e^{\tilde S^\dagger} 
e^{\tilde S}\prod_{n=1}^\infty \prod_\mu\prod_\nu \sum_{l=0}^\infty 
{(-)^l\over l!}\left(
A^\mu_nM_{\mu\nu}B^\nu_n\right)^l \vert N_m^\alpha \vert \prod_{k}^\infty
\prod_\rho\prod_\sigma 
\sum_{s=0}^\infty {(-)^s\over s!}\left(A^{\rho\dagger}_kM_{\rho\sigma}
B^{\sigma\dagger}_k\right)^s\Bigg\vert 0 
(\theta)\right\rangle\!\!\!\right\rangle ,
\label{Aminter}
\ee
where $\tilde{S}$ represents the exponential operator for the thermal tilde 
part of the brane state and $B=N_{p}\left( F,\theta \right)
\delta ^{\left( d_{\perp}\right) }\left( \hat{q}-y\right)$. 
By expanding the products in (\ref{Aminter}), one
is left with the expression of ${\cal A}_m$ in terms of states that describe
the number of excitations of string in each direction of space-time and 
for each oscillation mode. These states are orthogonal and of unit norm and
after some short but tedious algebra one can show that the (\ref{Am}) has the
following form  
\be
B^2\tilde B^2
\sum_{t_1^{1,1},\cdots, t_\n^{24,24}}
\sum_{s_1^{1,1}\cdots s_\n^{24,24}}
\sum_\rho\sum_\sigma 
(M_{24,24})^{^{2t_1^{24,24}}}\cdots 
(M_{1,1})^{^{2t_\n^{1,1}}} 
(M_{24,24})^{^{2s_1^{24,24}}}
\cdots (M_{1,1})^{^{2s_\n^{1,1}}}s_m^{\rho,\sigma}.
\label{finalAm}
\ee
Here, $s_i^{\rho,\sigma} = 0,1,\cdots, \infty$ \quad, $i=1,2,\cdots, n$ 
\quad, $n\to \infty$ \quad $\rho, \sigma = 1, \cdots , 24$ represent the 
indices for all excitation $s$ of all frequencies $n$ and in all space-time
directions. The expression (\ref{finalAm}) is not normalized. It contains the
full dependence on the Kalb-Ramond field in the entropy (\ref{rightentro}).
The temperature dependence of entropy is contained only in the $n_k$ terms,
more exactly in $\l$'s. However, it is not allways possible to write
down the explicit form of $\l$`s as functions of temperature even in the 
case of a single generator. (The relation leading to this function is 
obtained by equating the corresponding coefficient in the Bogoliubov 
transformation to the statistical distribution in the thermal vacuum 
\cite{tu}.) In this case, approximation or numerical methods should be used.   
The total entropy of the $D$-brane is obtained as twice the (\ref{rightentro})
since the left modes contribute with the same ammount to it.

\section{Final Discussions and Conclusions}

To conclude, we have analysed the $SU(1,1)$ thermal group formed by all
possible unitary thermal Bogoliubov generators in the case of the bosonic 
string and $D$-brane. The reason for this analysis is that we have 
constructed the thermal brane and string in the framework of TFD where 
$SU(1,1)$ represents the most general structure underlying the Bogoliubov 
transformations. By choosing a certain type of parameters $\theta$ the
most general Bogoliubov transformation can be fixed to be unitary or 
non-unitary. We have choosen a unitary trasformation in order to preserve the
structure of the Hilbert space at zero temperature and the usual 
interpretation of quantum mechanics. However, the tilde invariance of the 
thermal vacuum of the theory is lost with this choice, which is an undesirable
feature in TFD. The solution is to construct the thermal vacuum as the direct
product between the original thermal vacuum, i. e. the vacuum obtained by 
acting with the Bogoliubov transformation on the vacuum at zero temperature,
and its conjugate under the tilde operation. In this case, the thermal vacuum
factorizes in a tilde and a non-tilde part as does the vacuum at zero 
temperature. By using a non-unitary transformation, one would have lost 
states from the Fock space and the isomorphism between the Fock space and
its dual conjugate. The normalization constant of the $D$-brane state
would have changed and we would have expected that different bra coherent 
states
satisfied the boundary conditions. With two non-isomorphic bra and ket states
it would not have been possible to take the average of the entropy operator in
one $D$-brane state. Note that the general Bogoliubov transformation be
simultaneously unitary and generated by a tilde invariant Bogoliubov operator
implies that two generators of $SU(1,1)$ do not appear in it. Therefore, we
might conclude that unless the thermal vacuum is the product of the original 
thermal vacuum with its tilde conjugate, there is no $SU(1,1)$ thermal group
whose general transformation be compatible with the unitarity of quantum 
mechanics and the tilde invariance of TFD.
 
We have obtained the entropy of the closed string in (\ref{closedstringentr})
and the entropy of the bosonic $D$-brane as twice the entropy of the
right-modes (\ref{rightentro}). An analysis of this expression in various 
temperature limits should
be performed in an approximation or a numerical scheme for the $\theta$'s.
However, if one supposes that the $\theta$'s are monotonous functions 
at least at low temperatures, then, by taking $\l_{ik} \sim \e$ 
for $i=1,2,3$, we see that as $\e \rightarrow 0$ the entropy goes as
$$
\sum_m({\cal A}_m + \d_{mm}{\mbox Tr}\delta^{\m \n}),
$$ 
which in a normalized theory should be a finite constant. This represents an
improved entropy at low temperatures compared with the entropy given in 
(\cite{agv1}) which is divergent in that limit.

Note that the method used for computing the $D$-brane entropy in 
(\cite{agv1}) and in the present paper are slightly different from the 
standard TFD.
Indeed, since the $D$-branes are states in the Fock space of the bosonic string
vacuum different from the vacuum state of the conformal field theory, the 
entropy of $D$-brane was identified with the entropy of the closed string
in this coherent boundary state. A standard TFD would require to identify the
very $D$-brane with some vacuum state in a field theory, but no such of 
approach to $D$-branes is known at present. Also, as noted in \cite{cei}, 
our approach to thermal strings is different from the one in literature
in which understanding the ideal gase of the bosonic strings and the string 
field theory were the main motivations to implement TFD in ensembles of strings
and in string field theory \cite{yl1,yl2,yl3,fn1,hf,fn2}
much as was done for the standard field theory.
In our approach, due to the interpretation of brane, the TFD has been 
applied to the bosonic vacuum of string theory, i.e. to the two dimensional 
conformal field theory describing it, rather than to ensembles of strings
or string fields. Therefore, we have been working with the thermal bosonic 
vacuum and its thermal fluctuations some of which are thermal $D$-branes.

Let us end by observing that we have considered just a single solution of
$D$-brane type and not all solutions obtained in \cite{ivv,agv1}.
The reason for that is that the other solutions appear when the vacuum is
not invariant under tilde operation. If we impose the tilde invariance as
is done in TFD, the degeneracy of $D$-brane solutions should be removed. 

{\bf Acknowledgements}
We would like to thank B. C. Vallilo, P. K. Panda, D. Nedel,
B. M. Pimentel, H. Q. Placido and W. P. de Souza for useful discussions. 
I. V. V. also acknowledge to S. A. Dias and 
J. A. Helay\"{e}l-Neto for hospitality at DCP-CBPF and GFT-UCP 
where part of this work was done. A. L. G. acknowledges a CAPES doctoral 
fellowship. I. V. V. was supported by a FAPESP postdoc 
fellowship.


\newpage
\begin{thebibliography}{99}

\bibitem{mvm} M. V. Mozo, Phys. Lett. B{\bf 388}(1996)494 

\bibitem{mvmlfb}J. L. F. Barbon and M. V. Mozo, Nucl. Phys. B{\bf 497}(1997)236

\bibitem{rab1} S. A. Abel, J. L. F. Barbon, I. I. Kogan, E. Rabinovici,

JHEP 9904(1999)015

\bibitem{rab2} J. L. F. Barbon, E. Rabinovici, JHEP 0106(2001)029

\bibitem{ek} E. Kiritsis, JHEP 9910(1999)010

\bibitem{ekwt} E. Kiritsis and W. Taylor, hep-th/9906048

\bibitem{kog1} G. Dvali, I. I. Kogan and M. Shifman, Phys. Rev.

D{\bf 62}(2000)106001

\bibitem{kog2} S. Abel, K. Freese and I. I. Kogan, JHEP 0101(2001)039

\bibitem{kog3} I. I. Kogan, A. Kovner and M. Schvellinger, hep-th/0103235

\bibitem{od} A. Bytsenko, S. Odintsov and L. Granada, Mod. Phys. Lett.

A{\bf 11}(1996)2525

\bibitem{rsz} J. Ambjorn, Yu. Makeenko, G. W. Semenoff and R. Szabo,

Phys. Rev. D{\bf 60}(1999)106009

\bibitem{gub} S. S. Gubser, I. R. Klebanov, M. Rangamani, E. Witten,

hep-th/0009140


\bibitem{as} A. Sen, Nucl. Phys. B{\bf 440}(1995)421

\bibitem{sv} A. Strominger and C. Vafa, Phys. Lett. B{\bf 379}(1996)99

\bibitem{ls} L. Susskind, hep-th/9309145

\bibitem{hp} G. Horowitz and J. Polchinski, Phys. Rev. D{\bf 55}(1997)6189

\bibitem{bfks1} T. Banks, W. Fischler, I. Klebanov and L. Susskind, 

Phys. Rev. Lett. 80(1998)226, JHEP 9801(1998)008

\bibitem{cm}S. Chaudhuri and D. Minic, Phys. Lett. B{\bf 433}(1998)301

\bibitem{ks} I. Klebanov and L. Susskind, Phys. Lett. B{\bf 416}(1998)62

\bibitem{eh} E. Halyo, hep-th/9709225

\bibitem{dmkr} S. Das, S. Mathur, S. Kalyana Rama and P. Ramadevi,

Nucl. Phys. B{\bf 527}(1998)187

\bibitem{hm} G. Horowitz and E. Martinec, Phys. Rev. D{\bf 57}(1998)4935

\bibitem{ml} M. Li, JHEP 9801(1998)009; M. Li and E. Martinec, hep-th/9801070

\bibitem{bfk} T. Banks, W. Fischler and I. Klebanov, Phys. Lett. B{\bf 423}(1998)54

\bibitem{lt} H. Liu and A. Tseytlin, JHEP 9801(1998)010

\bibitem{dm} D. Minic, hep-th/9712202

\bibitem{ast} A. Strominger, Phys. Rev. Lett. 71(1993)3397

\bibitem{ivvol} I. V. Volovich, hep-th/9608137

\bibitem{ms} J. Maldacena and A. Strominger, JHEP 9807(1998)013

\bibitem{mlmjpp} M. L. Meana and J. P. Pe\~{n}alba, Nucl. Phys. B{\bf 560}(1999)154-180

\bibitem{sm} S. Mukohyama, Mod. Phys. Lett. A{\bf 11}(1996)3035


\bibitem{clny} C. G. Callan, C. Lovelace, C. R. Nappi and S. A. Yost,

Nuc. Phys. B{\bf 288}(1987)525; Nuc. Phys. B{\bf 293}(1987)83;

Nucl. Phys. B{\bf 308}(1988)221 

\bibitem{pc} J. Polchinski and Y. Cai, Nucl. Phys. B{\bf 296}(1988)91

\bibitem{gw}M. B. Green and P. Wai, Nucl. Phys. B{\bf 431}(1994)131 

\bibitem{bg}M. B. Green, Nucl. Phys. B{\bf 381}(1992)201 

\bibitem{gg}M. B. Green and M. Gutperle, Nucl.Phys. B{\bf 476}(1996)484 

\bibitem{ml1} M. Li, Nucl. Phys. B {\bf 460}(1996)351

\bibitem{dgk}
U.~H.~Danielsson, A.~Guijosa and M.~Kruczenski,
JHEP {\bf 0109}, 011 (2001)
[arXiv:hep-th/0106201].

\bibitem{ck} C. Callan Jr. and I. Klebanov, Nucl. Phys. B{\bf 465}(1996)473 

\bibitem{dv1} P. Di Vecchia  and A. Liccardo, {\it D-Branes in String Theory},

hep-th/9912161, hep-th/9912275 

\bibitem{dv2} P. Di Vecchia, M. Frau, A. Lerda and A. Liccardo, Nucl. Phys.

B{\bf 565}(2000)397



\bibitem{ivv} I. V. Vancea, Phys. Lett. B487(2000)175



\bibitem{agv} M. B. C. Abdalla, A. L. Gadelha and I. V. Vancea, 

Phys. Lett. A{\bf 273}(2000)235



\bibitem{agv1} M. B. C. Abdalla, A. L. Gadelha and I. V. Vancea,

Phys. Rev. D{\bf 64}(2001)086005 


\bibitem{cei} M. C. B. Abdalla, E. Graca and I. V. Vancea, 
M. C. Abdalla, E. L. Graca and I. V. Vancea,
Phys.\ Lett.\ B {\bf 536}, 114 (2002)

\bibitem{tu} Y. Takahashi and H. Umezawa, Collective Phenom. 2(1975)55

\bibitem{um} H. Umezawa, {\em Advanced Field Theory: Micro, Macro and

Thermal Field} (American Institute of Physics, 1993)

\bibitem{cu} H. Chu and H. Umezawa, Int. J. Mod. Phys. A{\bf 9} (1994) 2363.

\bibitem{gsw} M. B. Green, J. H. Shwarz and E. Witten, {\em Superstring

Theory} (Cambridge University Press 1987).

\bibitem{cha} S. Chaturvedi and V. Srinivasan, J. Phys. {\em A32} (1999)1909.

\bibitem{eber} K. W\'{o}dkiewicz and J. H. Eberly, J. Opt. Soc. Am. {\bf B3}

(1985) 485.

\bibitem{hua} K. Huang, {\em Statistical Mechanics} (John Wiley \& Sons,

1987). 

\bibitem{yl1} Y. Leblanc, Phys. Rev. D{\bf 38}(1988)3087

\bibitem{yl2} Y. Leblanc, Phys. Rev. D{\bf 36}(1987)1780; Phys. Rev. 
D{\bf 37}(1988)1547; 
Phys. Rev. D{\bf 39}(1989)1139; Phys. Rev. D{\bf 39} (1989)3731

\bibitem{yl3} Y. Leblanc, M. Knecht and J. C. Wallet, Phys. Lett. B{\bf 237}
(1990)357

\bibitem{fn1} H. Fujisaki and K. Nakagawa, Prog. Theor. Phys. 
{\bf 82}(1989)236; 
Prog. Theor. Phys. {\bf 82}(1989)1017; Prog. Theor. Phys. {\bf 83}(1990)18; 
Europhys. Lett. {\bf 20}(1992)677; Europhys. Lett. {\bf 28}(1994)471 

\bibitem{hf} H. Fujisaki, Il Nuovo Cimento, {\bf 108}A(1995)1079

\bibitem{fn2} H. Fujisaki and K. Nakagawa, Europhys. Lett. {\bf 35}(1996)493

\end{thebibliography}
\end{document}